# BEAM STABILIZATION AND INSTRUMENTATION SYSTEMS BASED ON INTERNET OF THINGS TECHNOLOGY

D. P. McGinnis†, MaxIV, Lund, Sweden

## *Abstract*

A longitudinal coupled bunch mode diagnostic was built to identify and characterize coupled bunch modes in the Max IV 3 GeV ring. A longitudinal coupled bunch Mode 0 dipole instability was identified and a longitudinal Mode 0 dipole damper was built which cured the instability. As a separate project, a radio frequency current transformer was built as an emergency backup to the DC current transformer. All the devices where built using the open-source Blinky-Lite Internet of Things platform that enabled quick and inexpensive deployment.

## STATEMENT OF THE PROBLEM

MaxIV possesses a 3 GeV synchrotron light source operating at an RF frequency of 100 MHz with enough installed power for over 250 mA of beam current. The Max IV 3 GeV ring is designed to operate in the long bunch mode (bunch lengths > 500 ps rms) using passive third harmonic Landau cavities.

As of April 2018, it was not possible to operate in long bunch mode for beam currents exceeding 200 mA because of longitudinal instabilities. The Max IV 3 GeV ring does possess a longitudinal bunch-by-bunch feedback system but the system was not effective in the long bunch mode. It was surmised that the longitudinal instability was a dipole mode 0 coupled bunch mode instability but there were no diagnostics to definitively prove this assertion.

## STEPS TO SOLUTION

First, build a coupled bunch mode analyser diagnostic to determine which coupled bunch mode instabilities are present and the growth rate of these modes. Then after the offending mode has been identified, build narrow band but high gain coupled bunch feedback systems to damp offending modes. However, both of these steps are fairly significant tasks requiring time and resources. However, to prove MaxIV's ground breaking approach of using multi-bend achromats [1] a success, it was crucial that this problem be solved quickly.

However, MaxIV is a small lab (~200 people) with limited human and financial resources. There are limited software and IT resources available because delays in beamline construction has absorbed the most of the control system resources. In addition, since MaxIV employs few technicians, there are limited technical resources with electrical engineers laying-out and building their own electronics boards.

For these reasons, the coupled bunch mode diagnostic system and the beam damping system were built using the open-source Blinky-Lite [2] platform which will be discussed later in the paper.

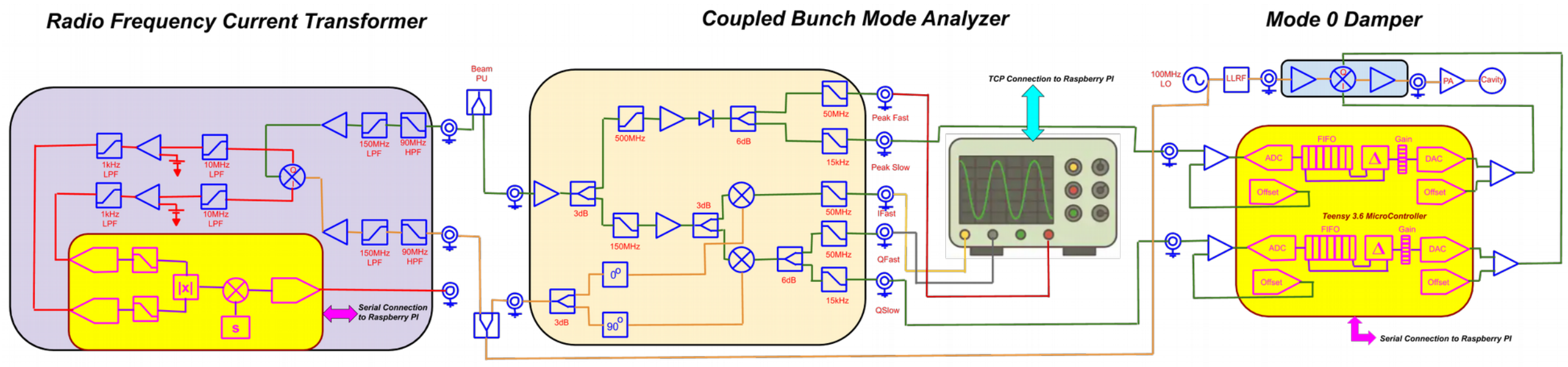

Figure 1: Coupled Bunch Mode analyser front end

## *Bunch by Bunch Front End*

Figure 1 shows the radio frequency front end electronics for the couple bunch mode analyser. A beam signal from a BPM pickup is split into two with one signal high passed filtered at 500 MHz. The 500 MHz beam signal will be used to measure the bunch length. After the high pass filter it is amplified and then rectified with a standard RF diode. The rectified signal is further split into two with a wideband resistive splitter with one signal low pass filtered at 50 MHz which is the one half the RF frequency in the Max IV 3 GeV ring. This signal will be digitized with an oscilloscope and compared to the 100 MHz beam signal to calculate the bunch-by-bunch bunch length. The other rectified signal is filtered at 15 kHz signal and will be used as a input for a Mode 0 quadrupole damper. The 15 kHz frequency was chosen because it is over 15 times larger than the synchrotron frequency.

The other side of the beam signal is low pass filtered at 150 MHz. The 150 MHz frequency was chosen to remove the bunch length dependence on the amplitude of the signal while preserving the bunch-by-bunch content. The signal is then amplified and separated into the in-phase and quadrature-phase components by splitting the signal to two RF mixers in which the LO port of one of mixer is fed with the in-phase component of the 100 MHz master oscillator and the other mixer LO port is fed with the quadrature phase of the master RF oscillator.

† david.mcginnis@maxiv.lu.se

The IF signals from both mixers are filtered with a low pass filters of 50 MHz to preserve the bunch-by-bunch information. Then both signals are digitized with an oscilloscope. Before digitization, the quadrature phase signal is further split into two with a wideband resistive splitter and one of the signals is further filtered down to 15 kHz to be used in a Mode 0 dipole damper.

The oscilloscope is interfaced with a Raspberry Pi 3B+ linux computer over the ethernet port. The Raspberry Pi 3B+ is the device-message adapter for a Blinky-Lite device which will be discussed later in this paper. A number of Blinky-Lite web applications [3][4][5] were developed to display the couple bunch mode spectrum. Figure 2 shows some screen shots of the web application for the coupled bunch mode diagnostic.

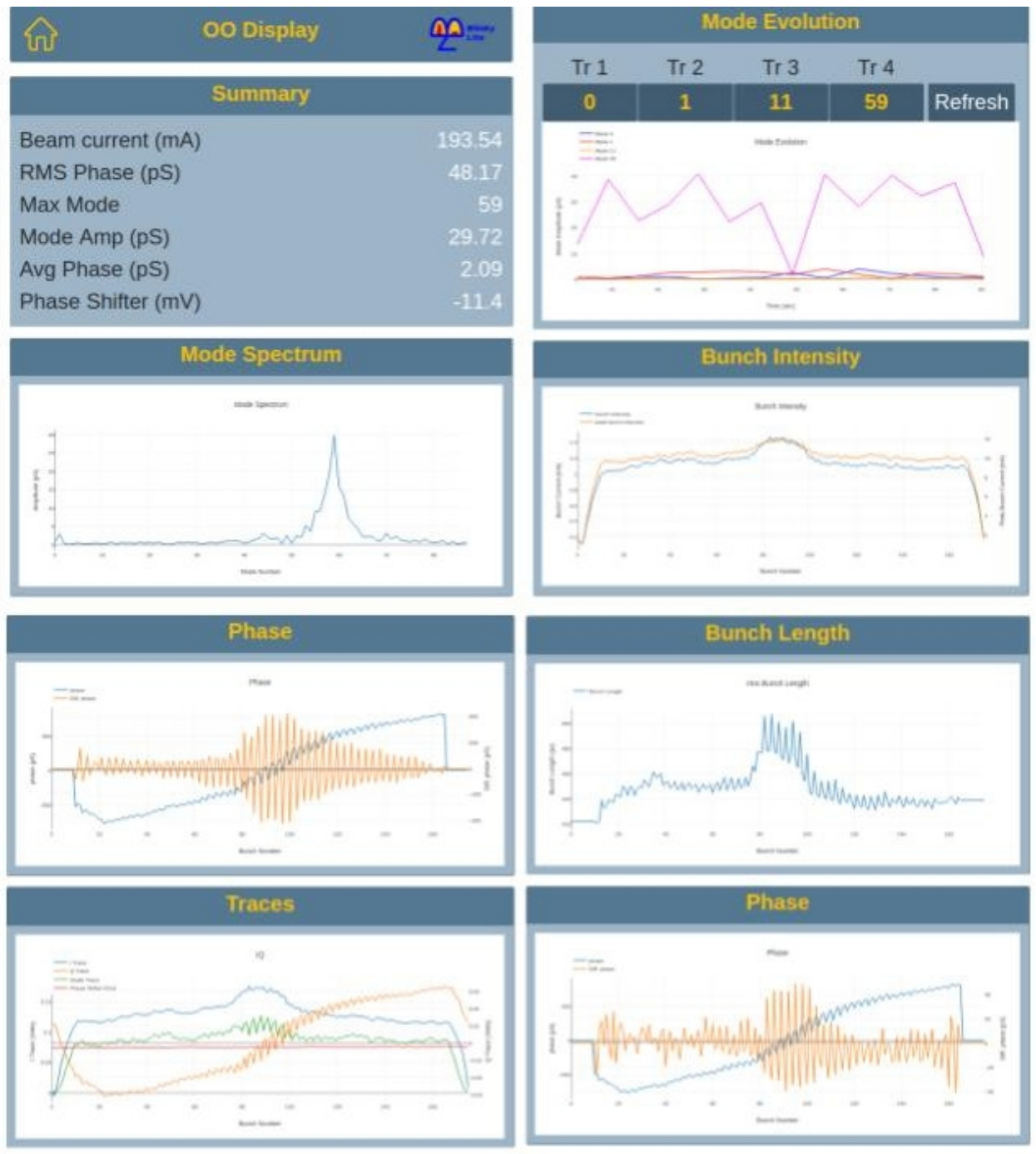

Figure 2: Oscillation Overthruster display for displaying the coupled bunch mode spectrum and associated waveforms.

## *Mode 0 Damper*

The coupled bunch mode diagnostic showed that the offending mode was the Mode 0 dipole mode. Mode 0 is where all bunches oscillate around the centre of their respect buckets with the same phase. A mode 0 dipole damper was constructed using a Teensy 3.6 [6] as shown in Figure 1. The Teensy 3.6 is a a 32 bit 180 MHz ARM Cortex-M4 processor with a floating point unit. The Teensy 3.6 contains a number of 12 bit ADCs that can acquire data at 100 kHz. In addition, there are two 12bit DACs that can convert at 100 kHz as well. The low frequency (15 kHz) quadrature signal described above was digitized by the Teensy 3.6. A programmable two tap finite impulse response filter takes the derivative of the phase motion to determine the sign of the energy kick to apply. A gain stage was implemented by bit shifting. A limiting function was also implemented. The differentiated signal was converted to analog by one of the Teensy 3.6 DACs. To phase shift the cavity fields with respect to the beam, the analog signal was applied to the Q channel of a 100 MHZ I-Q modulator placed in series with the RF drive to the cavities. In addition , provisions for a Mode 0 quadrupole damper was made by making a parallel channel in which the low frequency peak detector signal is digitized, differentiated with a two tap finite impulse response filter and applied to the I channel of the 100 MHZ I-Q modulator. The Teensy 3.6 was interfaced with a Blinky-Lite device-message-adapter implemented on a Raspberry Pi 3B through the serial ports of both devices. Control of the system was done through a Blinky-Lite application [7].

## *Timeline*

The project definition was given in early April 2018. The bunch-by-bunch front end and coupled bunch mode diagnostic were completed by the end of June 2018. The Mode 0 damper was commissioned in September of 2018. Figure 3 shows the effect of the Mode 0 damper on beam brightness. Long bunch operation for beam currents greater than 200 mA was achieved by October 2018. This rapid pace was only made possible by implementing the hardware in the Blinky-Lite control platform.

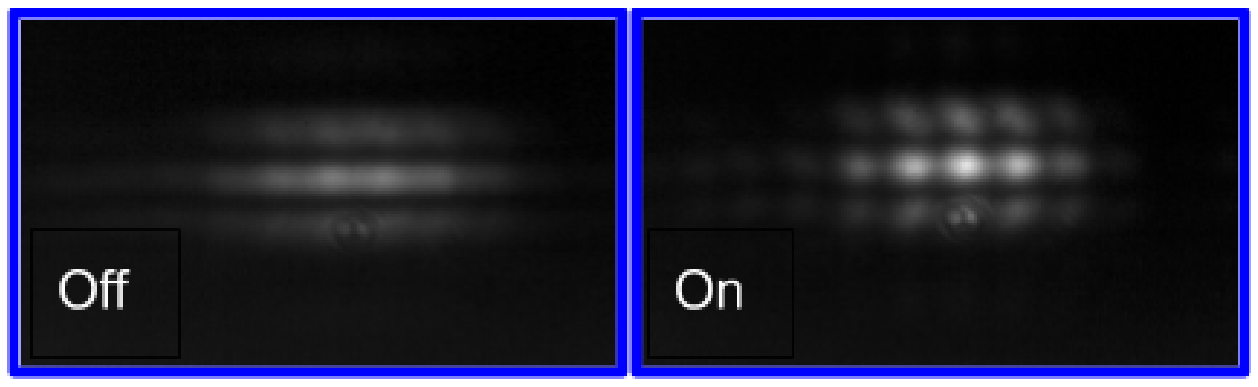

Figure 3. Max IV synchrotron light diagnostic beamline diffraction pattern with the Mode 0 damper off and on.

## *Radio Frequency Current Transformer*

At the end of the 2019 summer shutdown, the DC current monitor (DCCT) for the Max IV 3 GeV ring had a vacuum leak. The estimate for repair was four weeks which would delay the Fall 2019 run significantly. It was requested that an RF Current Transformer (RFCT) be built as an emergency substitute for the DCCT.

The RFCT circuit diagram is shown on the left hand side of Figure 1. It consisted of feeding the beam signal low passed at 150 MHz and the 100 MHz master local oscillator signal into a 100 MHZ I-Q de-modulator to produce in-phase and quadrature phase base-band signals. The base-band signals were low pass filtered at 1 kHz and fed into a Teensy LC micro-controller configured as a Blinky-Lite device controller.

The signals were digitized and the magnitude of the signals computed and scaled. The device controller fed the magnitude information to an output DAC that could be interfaced with the Machine Protection System and the Personal Protection System as an analog signal. In addition, the device controller communicated with Raspberry Pi Zero acting as a Blinky-Lite device message adapter through the serial port so the beam current data could be displayed to users. The RFCT had a range of 250 mA, a resolution better than 0.2 mA, and a a remotely adjustable bandwidth of 0.1 to 1 kHz. The I-Q channel offsets and differential gain, bandwidth, beam current and output voltage calibration constants could all be remotely configured. Figure 4 shows the web application for the RFCT.

The RFCT complete with data logging, archiving, and alarming was completed in less than one week and was ready for the startup.

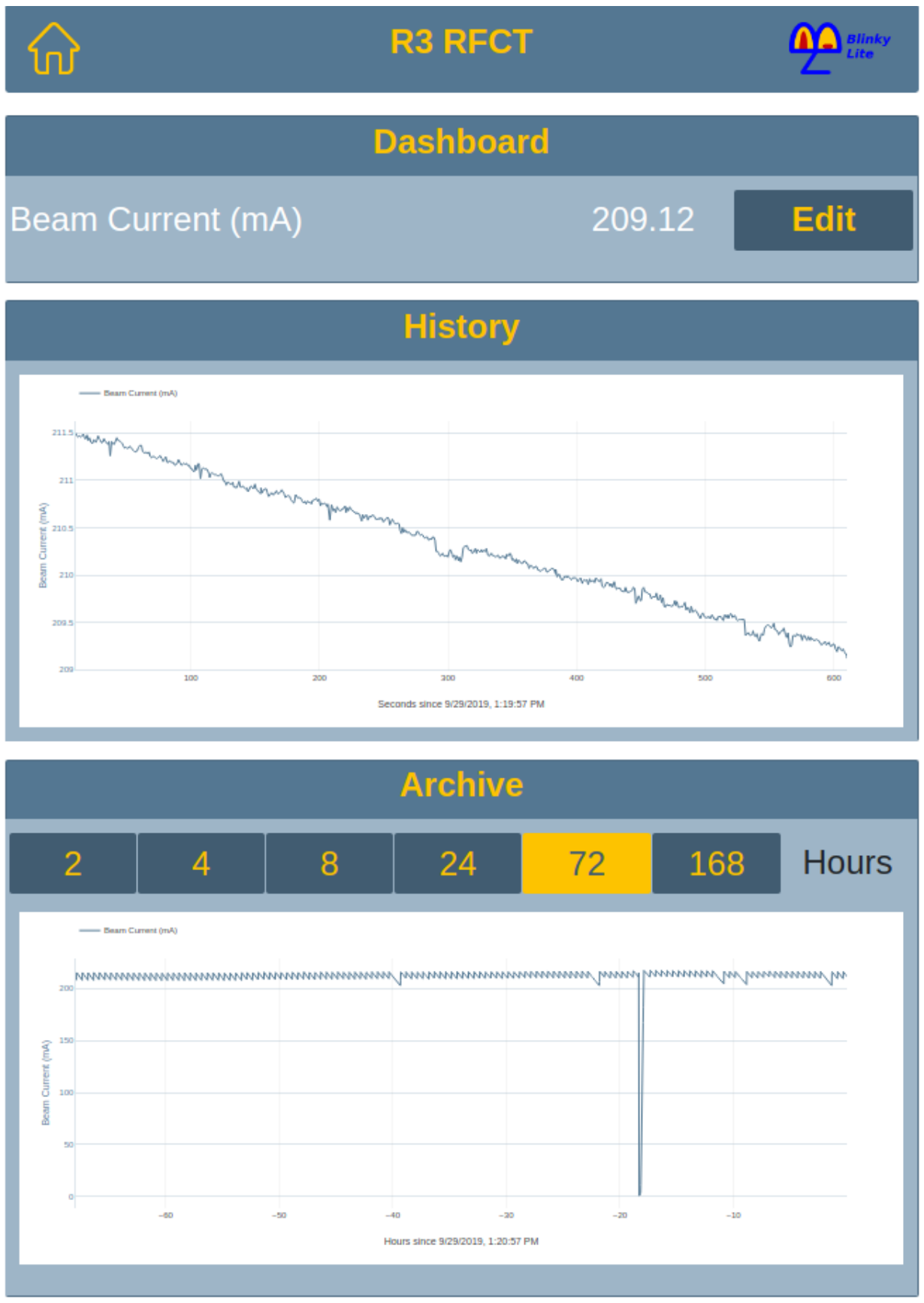

Figure 4. RFCT Web Application

## THE BLINKY-LITE CONTROL PLATFORM

Any instrumentation is useful only if the information supplied can be clearly displayed and manipulated by the user. In addition, almost all beam stabilization systems will have limited functionality if there is no user control and feedback at the outset of commissioning these systems. In today's big science laboratories, there has been a trend toward specialization with the controls experts having little knowledge of accelerator engineering and vice-versa for the accelerator engineers. This silo approach requires heavy project management oversight with the accompanying expense and long lead times.

To avoid the project management silos, the open-source Blinky-Lite control platform was designed to provide a control platform for engineers with minimal expertise programming control systems. In addition, with demands of high reliability for modern particle accelerators, access to the control system anywhere and at any time are an imperative. Thus, the Blinky-Lite control platform is based on Internet of Things (IoT) technology which is also simple to use and inexpensive.

Because it is based on high performance but inexpensive IoT computing placed close to the devices to control, the platform provides reliable and robust control. Applications are web-based giving control from anywhere in the world. The platform is flexible because it is completely open source for easy customization. Finally, the platform is designed for non-experts who have beginner knowledge in Javascript.

### *Features*

- **Cloud capable** - Cloud deployments give enhanced accessibility and deployment capability along with enhanced reliability and security (https:// and wss://)
- **Layered authentication** JSON Web Tokens for client-server transactions and authenticated MQTT broker for server-device transactions
- **JSON Device configuration** Flexible data types (scalar, vector, text, images, blobs,…) that are human readable and configurable
- **MQTT and Websocket communication** Publish-subscribe instead of polling protocols
- **SMS Alarm notification**
- **Graphical Node-Red code environment** Re-usable code and self documentation

### *Architecture*

The Blinky-Lite architecture is shown in Figure 5.

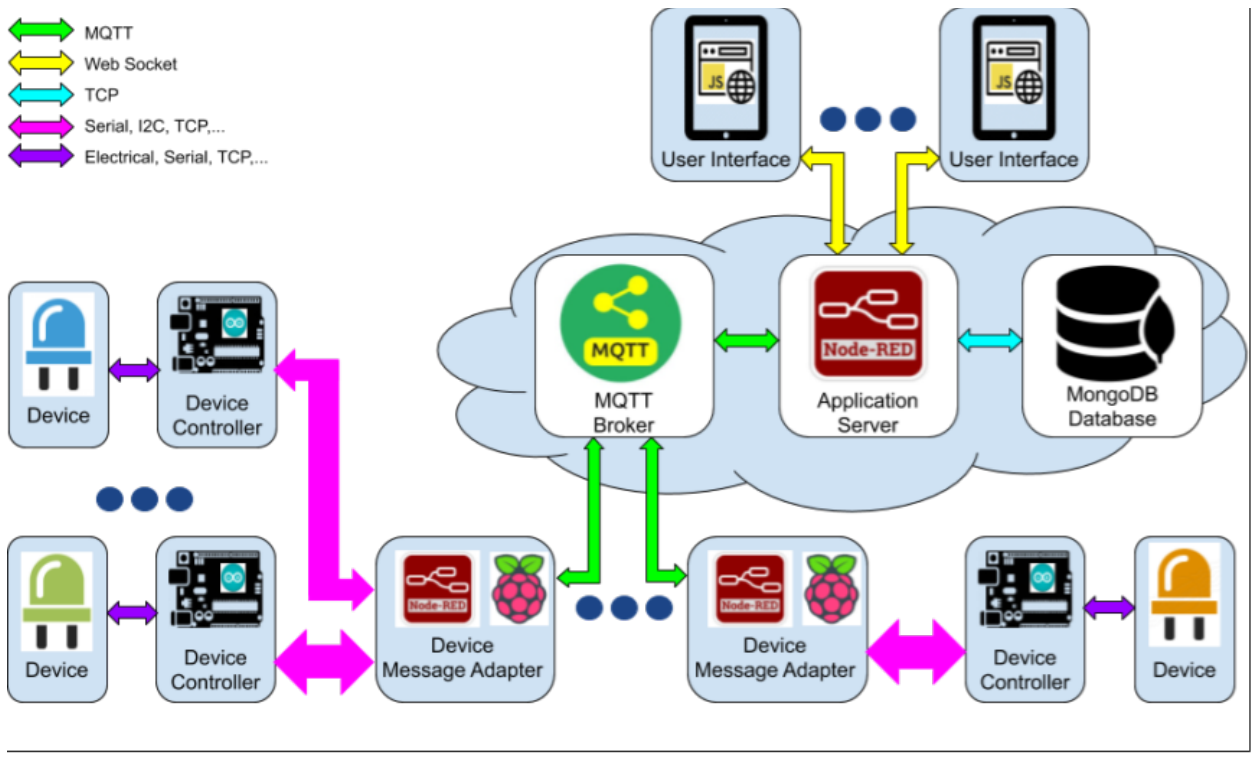

Figure 5. Blinky-Lite architecture

**Device** At the lowest level is the *device* which is some type a sensor or actuator. In the case of the Mode 0 damper, the sensor *device* is the down-converted beam signal and the actuator *device* is the RF phase shifter.

**Device Controller** Interfaced directly to the *device* is the *device controller*. There can be many devices connected to a single *device controller*. In many cases, the *device controller* is micro-controller. For example, in the Mode 0 damper, the *device controller* is the Teensy 3.6 in which the beam pickup interface is a 12 bit ADC and the RF phase shifter interface is a 12 bit DAC. As with many micro-controllers, the Teensy 3.6 is programmed in C using the Arduino IDE [8].

**Device Message Adapter** In turn, the *device controller* communicates with a *device message adapter* (DMA) usually through a serial port. A single DMA can communicate with a number of Device Controllers. The DMA is a linux computer, which in many cases, a Raspberry Pi 0 or 3B [9] is more than sufficient. In Blinky-Lite, the DMA is programmed using the Node-RED [10], a flow based, graphical programming environment for Node.js. The advantages of using Node-RED are many:

- There are a large ecosystem of nodes that drastically reduces the amount of coding required by the user.
- Flow based programming is a very natural way of organizing code in control system event drive applications.
- The programming environment is web-based allowing the user to program from anywhere in the world on almost any kind of platform.
- The graphical nature of the programming environment lends to natural self-documentation
- The code is stored in JSON [11] format which can be readily stored and tracked in a git [12] repository.

Figure 6 shows the Node-RED flow for the Max IV coupled bunch mode analyser.

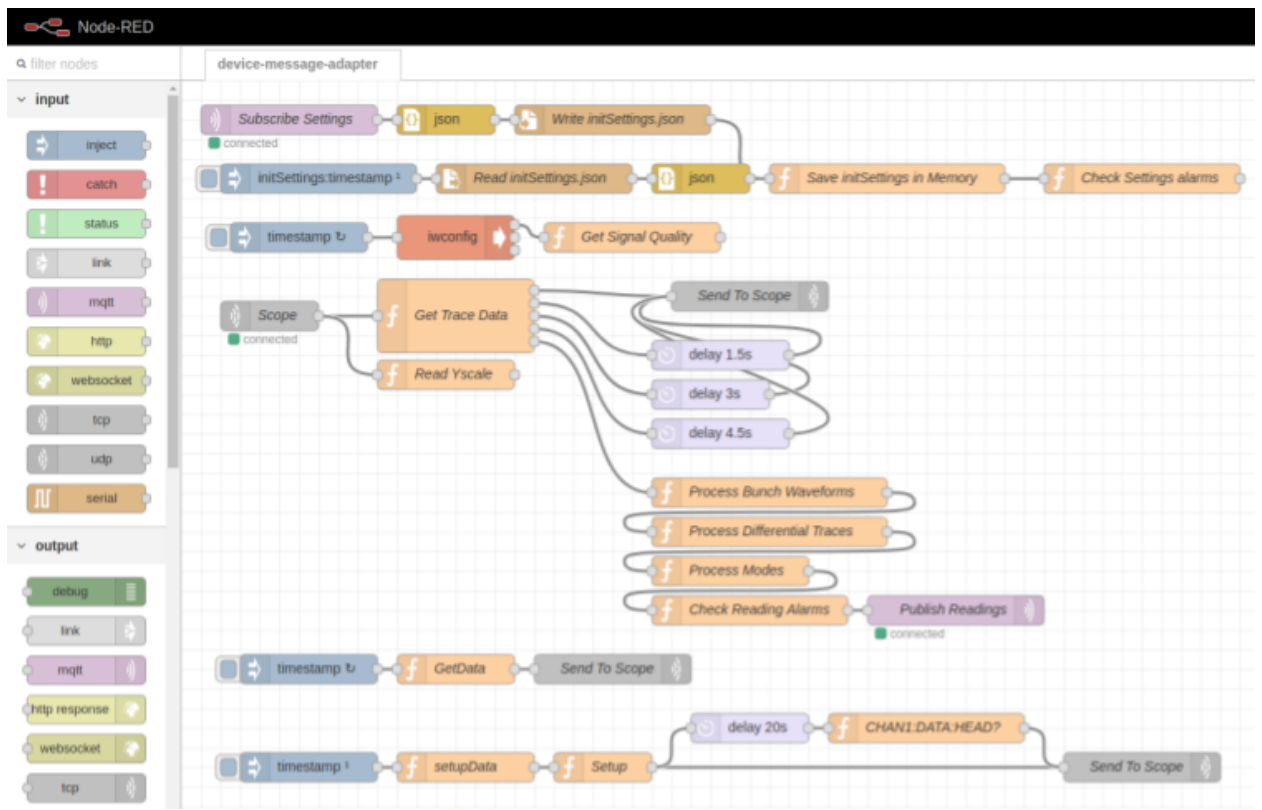

Figure 6. Node-RED flow for coupled bunch mode DMA

**MQTT Broker** Device data in a DMA is read and stored as a JSON object and is transmitted to an MQTT broker as the message payload. MQTT [13] is a machine-to-machine (M2M)/IoT connectivity protocol. It was designed as an extremely lightweight publish/subscribe messaging transport. It is useful for connections with remote locations where a small code footprint is required and/or network bandwidth is at a premium. The MQTT broker can be cloud-based which frees up the user from having to maintain the MQTT broker hardware and software. For example, the MQTT broker for the MaxIV Blinky-Lite applications is provided by shiftr.io [14]. The MaxIV MQTT network is shown in Figure 7. The MQTT broker receives messages from DMAs and transmits the messages to the Web Application Server.

**Web Application Server** The *Web Application Server* (WAS) collects and transmits data to DMAs via the MQTT broker. In addition, it archives the device messages into a database, handles user authentication, and device alarming. User authentication for setting devices is handled via JSON Web Tokens [15]. As with the MQTT broker, the WAS can be cloud-based freeing up the user from having to maintain the WAS hardware. The WAS also serves web-based user applications. Like the DMAs, in Blinky-Lite, the WAS is programmed using Node-RED. Figure 8 shows the Node-RED flow for the MaxIV custom Blinky-Lite web applications.

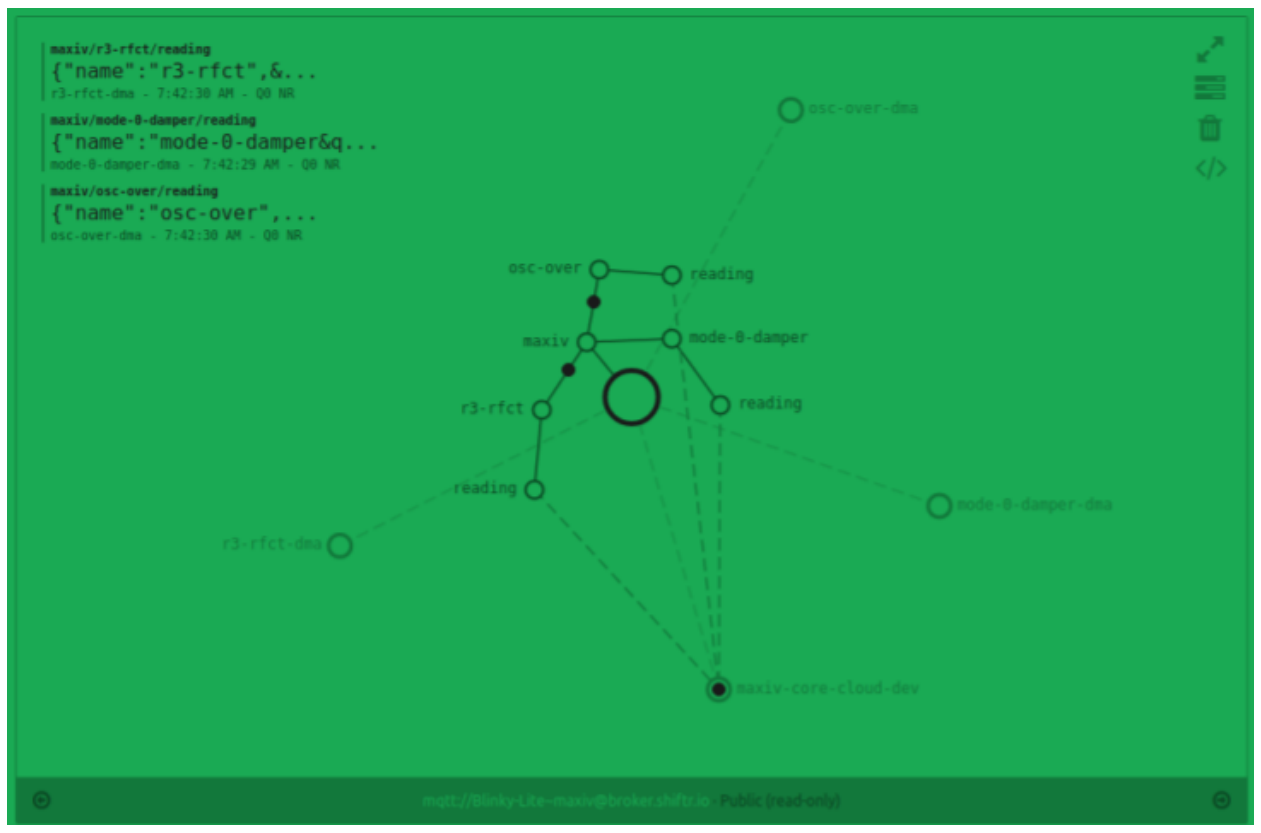

Figure 7. MQTT Network for MaxIV applications.

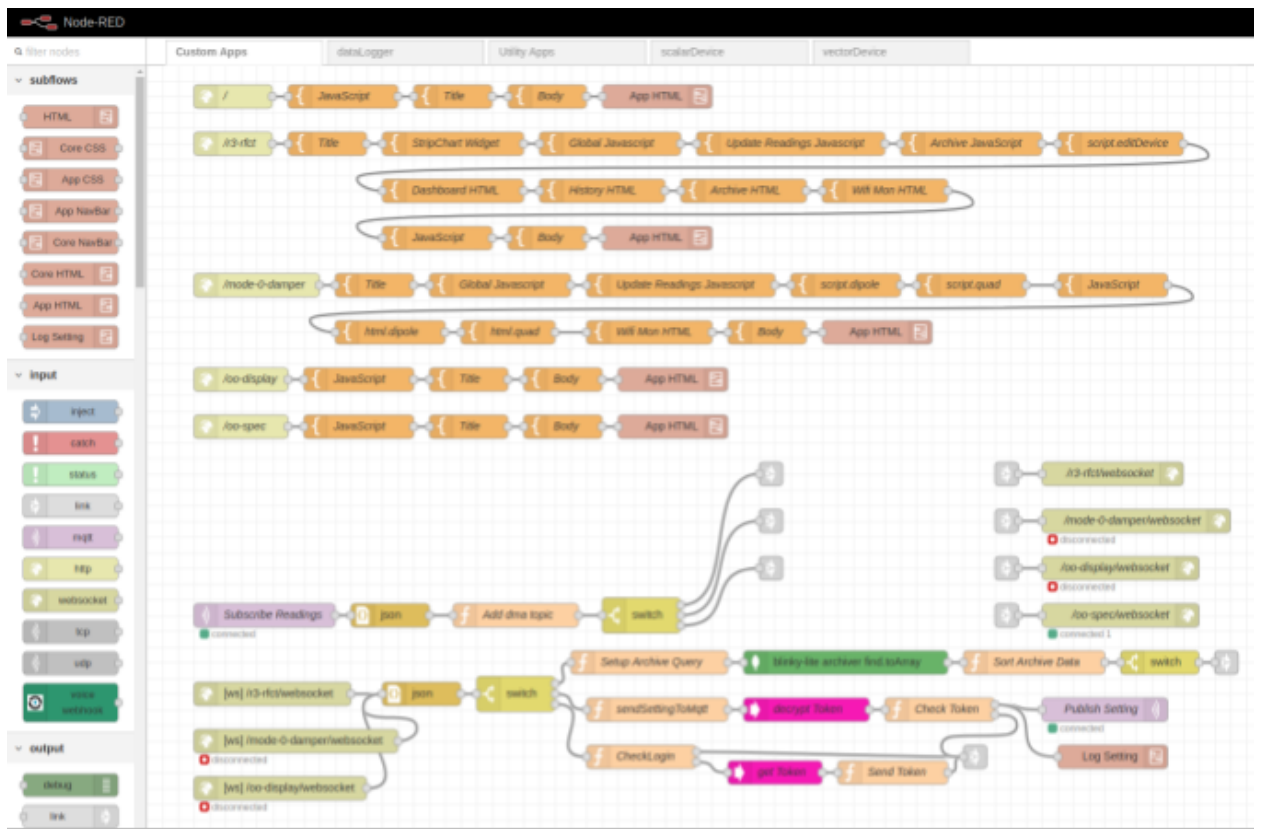

Figure 8. Custom web application flow for MaxIV

**Device Database** In Blinky-Lite, the device database is a MongoDB [16] database. The primary purpose of the database is for archival of data. MongoDB was chosen because the MongoDB records are JSON objects which matches well to the structure of the device data. Also MongoDB is non-relational which is easy to extend and define for a non-expert. As with the MQTT broker and the WAS, the MongoDB database can be cloud based.

**User Applications** In Blinky-Lite, user applications are web-based. Web-based applications give the user control from anywhere in the world on almost any hardware platform and deployment and maintenance of the applications is much more straightforward. Modern browsers today are powerful virtual machines that can serve as a platform for sophisticated client-side applications. In addition, there is a plethora of open-source third party software available for sophisticated displays and graphics. The client side application is programmed in JavaScript and HTML.

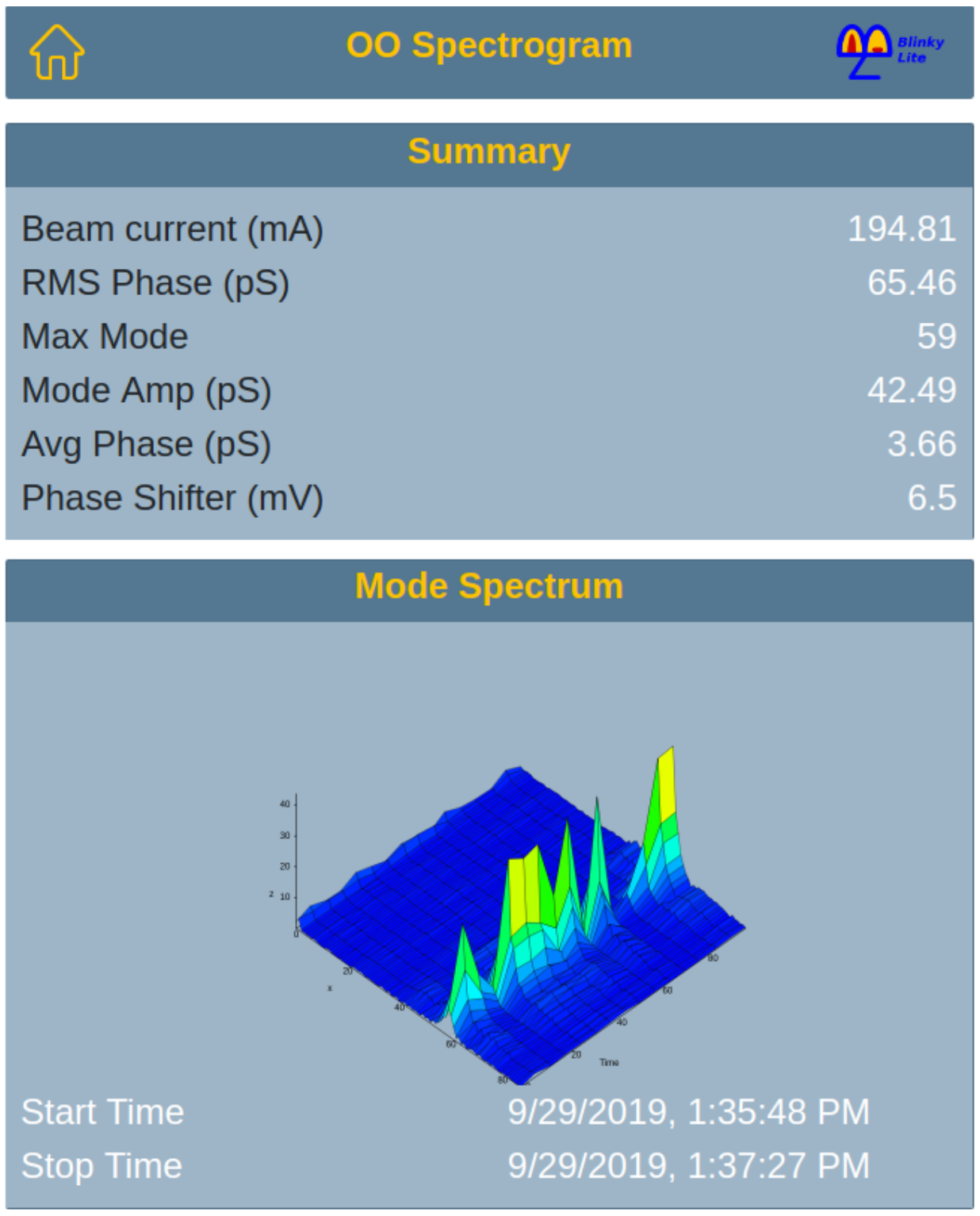

Figure 9. Max IV 3 GeV Ring coupled bunch mode spectrum

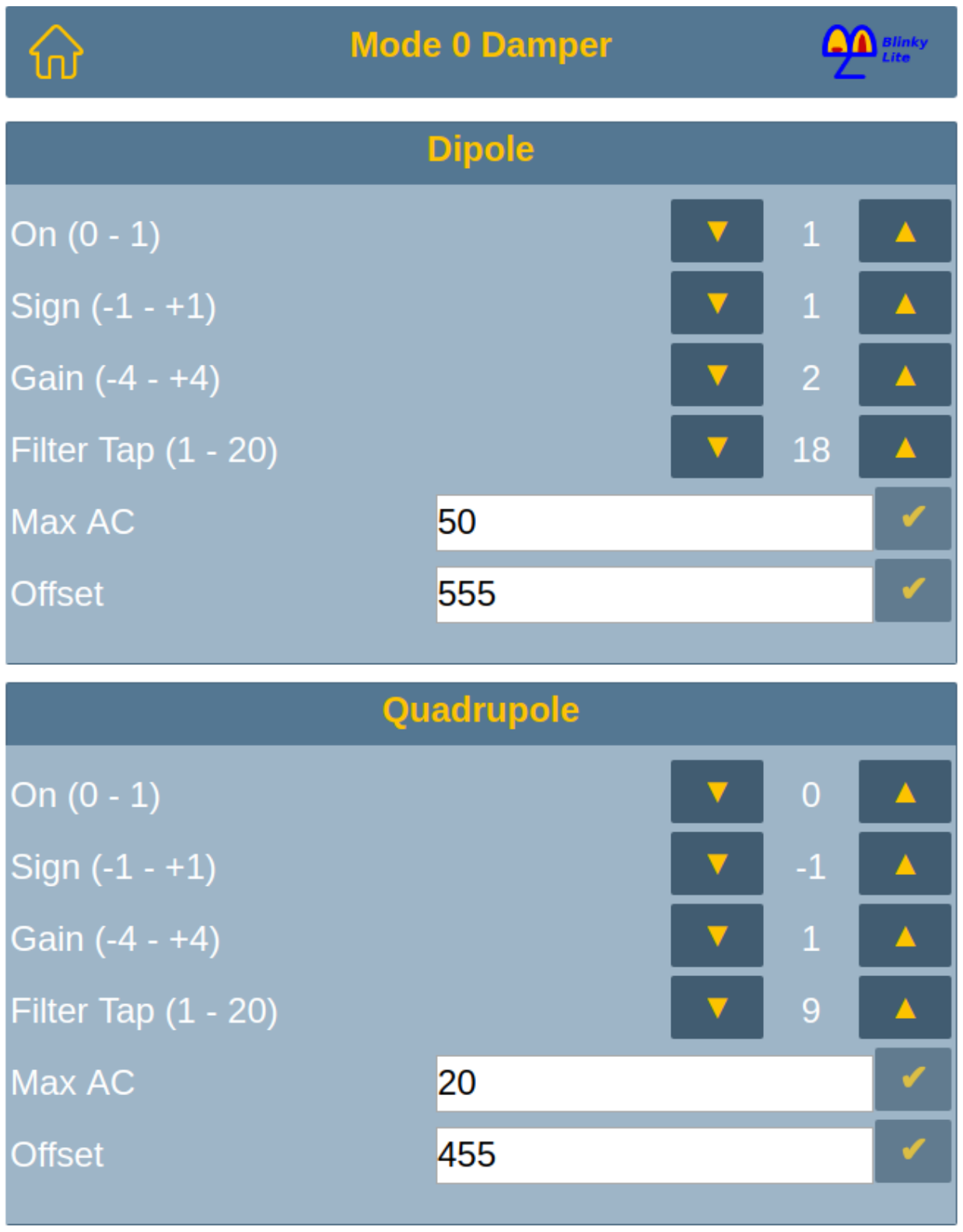

Figure 10. Mode 0 Damper control panel.

In addition to customized applications, Blinky-Lite comes with a suite of 8 core applications:

- Scalar plotter
- Scalar archive plotter
- Scalar alarm scanner (with sms notification)
- Scalar viewer for settings, and alarm limits
- Vector plotter
- Vector archive plotter
- Settings log
- Access log

## CONCLUSION

A longitudinal coupled bunch mode diagnostic instrument was built to identify and characterize coupled bunch modes in the Max IV 3 GeV ring. A longitudinal coupled bunch Mode 0 dipole instability was identified and a longitudinal Mode 0 dipole damper was built which cured the instability. As a separate project, a radio frequency current transformer was built as an emergency backup to the DC current transformer. All the devices where built using the open-source Blinky-Lite Internet of Things platform that enabled quick and inexpensive deployment.